\documentclass[manuscript,screen,nonacm]{acmart}
\acmConference[]{} 

\newcommand{\workshopname}{GenAICHI 2025: Generative AI and HCI at CHI 2025}
\newcommand{\licensedetails}{Licensed under a Creative Commons Attribution 4.0 International License (CC BY 4.0). Copyright remains with the author(s).}
\newcommand\extrafootertext[1]{
    \bgroup
    \renewcommand\thefootnote{\fnsymbol{footnote}}%
    \renewcommand\thempfootnote{\fnsymbol{mpfootnote}}%
    \footnotetext[0]{#1}%
    \egroup
}

\AtBeginDocument{ 
    \fancypagestyle{firstpagestyle}{
        \fancyhf{}
        \fancyfoot[L]{\sffamily\footnotesize \workshopname}%
        \fancyfoot[C]{\sffamily\footnotesize \thepage}
    }
    \fancyhf{}
    \fancyhead[L]{\sffamily\footnotesize\shorttitle}
    \fancyhead[R]{\sffamily\footnotesize\shortauthors}
    \fancyfoot[L]{\sffamily\footnotesize\workshopname}%
    \fancyfoot[C]{\sffamily\footnotesize\thepage}
    \extrafootertext{\licensedetails}
}
\usepackage{lipsum} 

\begin{document}

\title{ChatNekoHacker: Real-Time Fan Engagement with Conversational Agents}

\author{Takuya Sera}
\affiliation{%
  \institution{NEC Corporation, Neko Hacker}
  \city{Tokyo}
  \country{Japan}}

\author{Yusuke Hamano}
\affiliation{%
  \institution{NEC Corporation}
  \city{Tokyo}
  \country{Japan}}


\begin{teaserfigure}
  \centering
  \includegraphics[width=0.5\textwidth]{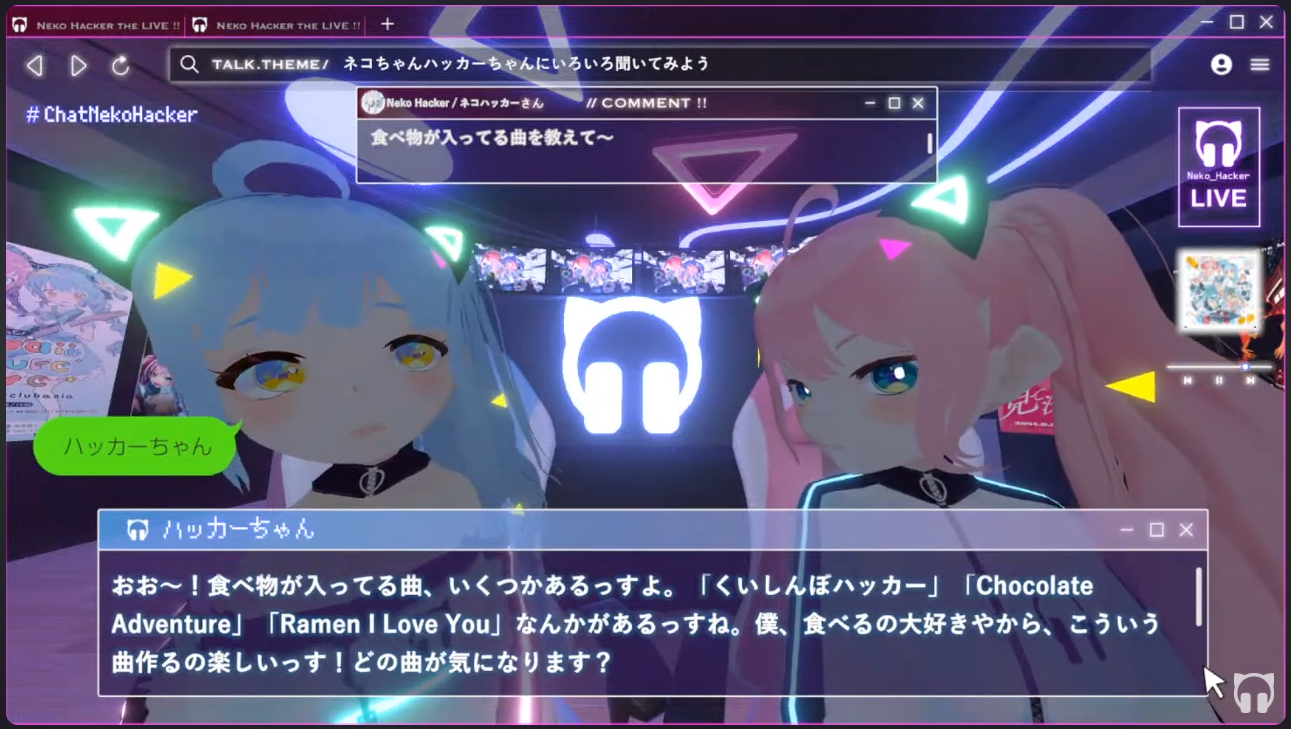}
  \caption{ChatNekoHacker User Interface}
  \label{fig:teaser}
\end{teaserfigure}

\begin{abstract}
ChatNekoHacker is a real‑time conversational agent system that strengthens fan engagement for musicians. It integrates Amazon Bedrock Agents for autonomous dialogue, Unity for immersive 3D livestream sets, and VOICEVOX for high quality Japanese text-to-speech, enabling two virtual personas to represent the music duo Neko Hacker. In a one‑hour YouTube Live with 30 participants, we evaluated the impact of the system. 

Regression analysis showed that agent interaction significantly elevated fan interest, with perceived fun as the dominant predictor. The participants also expressed a stronger intention to listen to the duo's music and attend future concerts. These findings highlight entertaining, interactive broadcasts as pivotal to cultivating fandom. Our work offers actionable insights for the deployment of conversational agents in entertainment while pointing to next steps: broader response diversity, lower latency, and tighter fact‑checking to curb potential misinformation.
\end{abstract}


\maketitle
\begin{figure*}
  \centering
  \includegraphics[width=1.0\textwidth]{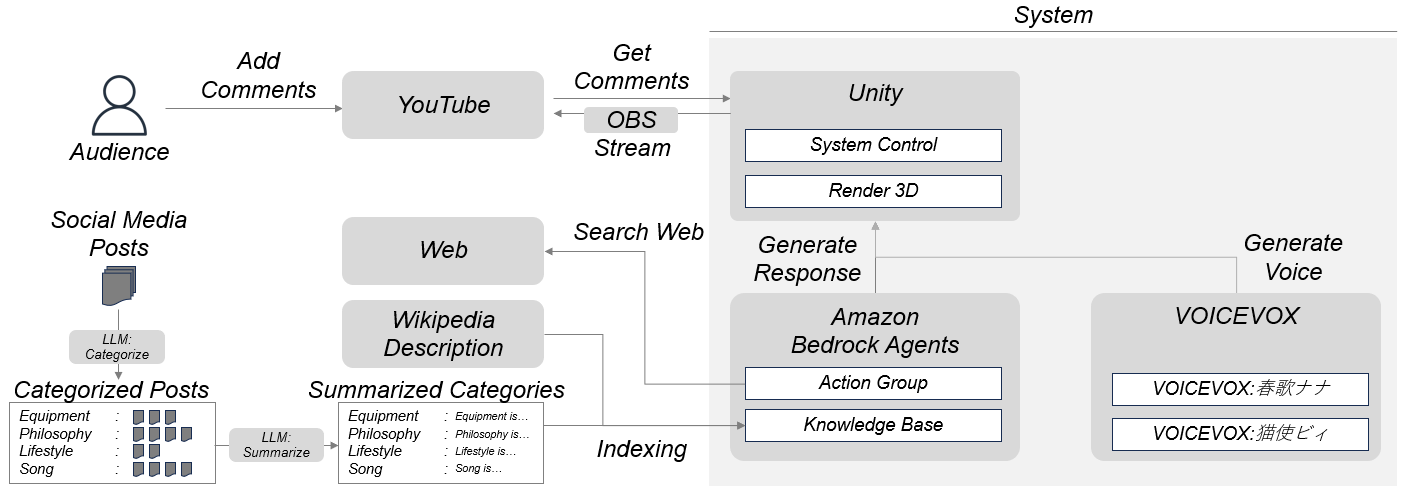}
  \caption{System Architecture}
  \Description{Diagram showing the architecture of the ChatNekoHacker system, including Unity, Amazon Bedrock Agents, and VOICEVOX components.}
  \label{fig:system_architecture}
\end{figure*}
\label{sec:system}
\section{Introduction}
In the entertainment domain particularly within the music industry the utilization of social media and dedicated applications for interactive engagement with fans has garnered significant attention as a key factor in enhancing brand value \cite{Edlom2021HangWM,Li2024BehaviorAI}. Maintaining fan interest and motivation to act is therefore of paramount importance.

Moreover, the rapid evolution of Large Language Models (LLMs) has introduced new possibilities in the design of conversational interfaces. Approaches employing LLM-based conversational agents---such as SimsChat \cite{Yang2024SimsChatAC} and Character-LLM \cite{Shao2023CharacterLLMAT}---have demonstrated that advanced finetuning can improve character consistency and factual accuracy. However, these methods typically require substantial computational resources and specialized expertise, presenting a significant barrier for adoption by artists and record labels.
In this paper, we present the development of a conversational agent based on a persona constructed from various information sources concerning the duo known as Neko Hacker, utilizing Amazon Bedrock Agents. Additionally, we built a system that automatically collects viewer comments from YouTube Live broadcasts and delivers responses through speech synthesis.

Figure~\ref{fig:teaser} shows a snapshot of the ChatNekoHacker system interface during a live broadcast.

During a one-hour live broadcast, surveys were collected from 30 viewers. The results were analyzed from the following two perspectives:
\begin{itemize}
  \item The contribution to enhancing fan interest and motivation to act.
  \item The advantages and challenges faced by artists when implementing fan engagement via a conversational agent.
\end{itemize}


\section{System}
Figure~\ref{fig:system_architecture} illustrates the overall system architecture.

\noindent 2.1 Unity

To recreate an engaging live broadcast experience similar to traditional human hosted sessions, a 3D character model and streaming environment were reproduced within Unity. Through Unity, the system facilitates the collection of comments from YouTube Live, response generation via Amazon Bedrock Agents, and voice synthesis using VOICEVOX.

\noindent 2.2 Amazon Bedrock Agents

Amazon Bedrock Agents provide functionalities for constructing autonomous agents. By leveraging a knowledge base, these agents can retrieve information from unstructured data sources and generate responses to user inputs. Moreover, through integrated action capabilities, they are able to execute functions such as web searches to obtain the latest information. In this study, by indexing summaries of group members' social media posts and activity overviews from Wikipedia into the knowledge base, agents can generate response based on past member statements. For each member, an LLM was used to classify the Social Media Posts into 15 different categories. Subsequently, a summary was generated for each category, and the summary results were indexed into the Knowledge Base.

Prompt engineering was used to control the system to speak in Kansai dialect, which is the Japanese dialect spoken by the members. Additionally, by implementing web search functionality within the action group, the system could integrate up-to-date information on live events and releases. Given that Neko Hacker is a duo, two distinct conversational agents---Neko-Chan and Hacker-Chan---were developed. 

\noindent 2.3 VOICEVOX

VOICEVOX \cite{VOICEVOX} is a text-to-speech and singing voice synthesis software compatible with Japanese. In the present study, the voice ``VOICEVOX: Haruka Nana'' was used for Neko-Chan, and ``VOICEVOX: Neko-tsuka Bii'' for Hacker-Chan.

\section{Experiments}
A one-hour live broadcast was conducted during which surveys were administered to 30 viewers; the results were subsequently analyzed. The survey items included the frequency with which respondents listened to Neko Hacker’s music, the number of times they had attended Neko Hacker’s live events, and a series of questions concerning their viewing experience and changes in their motivation. Responses were recorded on a five-point Likert scale (Agree, Slightly Agree, Undecided, Slightly Disagree, Disagree). The questions are presented in Table~\ref{tab:item_description}


According to the survey results, 83\% of respondents provided positive responses (Agree or Slightly Agree) to the item \emph{Interest: Increased interest in the artist}. To assess how interactions with the conversational agent contributed to this enhanced interest, a regression analysis using the least squares method was performed based on four survey items. The results are presented in Table~\ref{tab:stat_analysis}

\begin{table*}[ht]
  \centering
  \begin{minipage}[t]{0.48\textwidth}
    \centering
    \captionof{table}{Item and Description}
    \label{tab:item_description}
    \begin{tabular}{ll}
      \toprule
      Item       & Description(NH: Neko Hacker) \\
      \midrule
      Interest   & Increased interest in the artist \\
      Fun        & Conversation was enjoyable \\
      Useful     & Gained meaningful information \\
      Reality    & Made statements resembling the artist’s \\
      Unity      & Felt a sense of unity during broadcast \\
      Frequent   & Frequency of listening to NH’s music \\
      Experience & Number of times attended NH’s live events \\
      ListenMore & Felt like listening to more NH’s music \\
      JoinMore   & Felt like attending more NH’s live events \\
      Comments   & Free‑response comments \\
      \bottomrule
    \end{tabular}
  \end{minipage}\hfill
  \begin{minipage}[t]{0.48\textwidth}
    \centering
    \captionof{table}{Statistical Analysis Results}
    \label{tab:stat_analysis}
    \begin{tabular}{lccc}
      \toprule
      Item     & Positive Rate & Coefficient & p‑value \\
      \midrule
      Interest & 83\%           & —           & —       \\
      Fun      & 86\%           & 0.59        & 0.01    \\
      Useful   & 67\%           & 0.24        & 0.13    \\
      Reality  & 67\%           & 0.00        & 1.0     \\
      Unity    & 67\%           & 0.25        & 0.11    \\
      \bottomrule
    \end{tabular}
  \end{minipage}
\end{table*}


With an adjusted R-squared of 0.56, an F-statistic of 10.27, and a p-value of 0.00005 for the F-statistic, the regression model is both valid and statistically significant. The analysis revealed that ``Fun'' was the only factor contributing significantly to ``Interest,'' while ``Useful,'' ``Reality,'' and ``Unity'' had minimal influence and were not statistically significant. In free response comments, respondents noted positive aspects such as ``The responses were natural'' and ``The conversation closely resembled everyday dialogue.'' Conversely, comments highlighting areas for improvement---such as ``The conversation lacks variety'' and ``I expected more interesting responses''---indicate that viewers primarily sought an enjoyable experience.

\begin{figure*}[ht]
  \centering
  \includegraphics[width=1.0\textwidth]{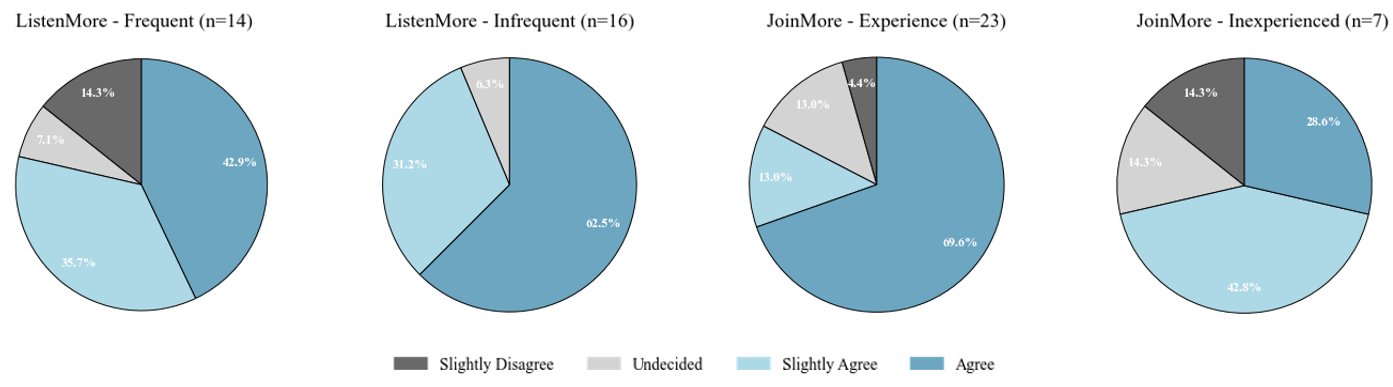}
  \caption{Results for ``ListenMore: Felt like Listening to More Songs'', ``JoinMore: Felt like Participating in More Live Events''}
  \Description{Bar chart comparing survey responses between Frequent and Infrequent groups regarding the willingness to listen to more songs.}
  \label{fig:graph}
\end{figure*}

Next, the contribution to enhancing the willingness to act was evaluated. Respondents who indicated that they listened to the music ``every day'' were categorized as the \emph{Frequent} group, while those who reported listening less than several times per week were categorized as the \emph{Infrequent} group. The response tendencies for the item \emph{ListenMore: Felt like listening to more songs} were compared between these groups. The results are presented in Figure~\ref{fig:graph}. Furthermore, respondents were categorized into two groups based on their live event participation: those who had attended live events were labeled as \emph{Experienced}, while those who had not were labeled as \emph{Inexperienced}. The response tendencies for the item \emph{JoinMore: Felt like participating in more live events} were then compared. A chi-square test conducted for the items \emph{ListenMore} and \emph{JoinMore} revealed no significant difference in the response distribution between the groups.

During the live broadcast, the following conversation took place. Although Japanese was used during the broadcast, an English translation is provided. "Spaceship" and "City Runner" are actual song titles, and the concept described by Hacker-Chan is based on the members' past posts.
\begin{itemize}
  \item Audience: Do you have a song or memory that has left the strongest impression on you during your production process?
  \item Hacker-Chan: That's a tough question to answer. But the song 'Spaceship' is particularly memorable. I created it with the image of a spaceship in mind, and it turned into a sparkling Vocaloid track. I also put a lot of care into the guitar phrasing. Additionally, 'City Runner' is another memorable piece. I made it with the idea of running through a futuristic city, and it gave off a feeling of being chased. I really enjoyed the production process.
\end{itemize}

\section{Discussion and Conclusion}
This study demonstrates that conversational agents have the potential to enhance fan engagement for music artists. In particular, the findings indicate that the enjoyment derived from the conversation is a critical factor in increasing interest in the artist, suggesting that highly entertaining interactions provided by conversational agents are key to boosting fan engagement. Moreover, the agents were found to also increase interest in both songs and live events, thereby potentially enhancing the willingness to act. An interesting observation was made during the live broadcast: a viewer commented regarding an out of stock item from an online store, and following the subsequent restocking, the item was purchased. This example indicates that real time conversational agents may foster purchasing intent and support consumer behavior.

However, this study is subject to several limitations. Due to the small sample size of 30 respondents, caution must be exercised when generalizing the results. Future work should address issues highlighted in the free response comments, such as enhancing the variety and interest of the responses, better incorporating conversational context during broadcasts, improving the interplay between the two conversational agents, and reducing the latency from comment collection to vocalization. Furthermore, some responses generated by the conversational agents included inaccuracies or chronologically inconsistent information, posing a risk of disseminating misinformation to viewers.

Looking ahead, there is potential for conversational agents to evolve into tools that more accurately reflect an artist’s personality and support deeper fan relationships. By more precisely incorporating an artist’s past statements, works, and musical style, it may be possible to facilitate conversations that are more authentically representative of the artist. Additionally, such agents could learn individual fan preferences and interests to provide tailored conversational experiences. With continued technological advancements, conversational agents hold the promise to transform fan engagement within the entertainment industry by establishing a novel communication channel between artists and their fans.

\begin{acks}
The authors would like to express their gratitude to all participants in the live broadcast and survey.  
The authors note that “NEC Corporation” denotes their institutional affiliation only, and that the research was conducted independently by Neko Hacker and the authors as individual researchers.
\end{acks}


\bibliographystyle{ACM-Reference-Format}
\bibliography{references}

\end{document}